\documentclass[%
reprint,
superscriptaddress,
showpacs,preprintnumbers,
amsmath,amssymb,
prl
]{revtex4-1}

\usepackage{graphicx}   %include figure files
\usepackage{dcolumn}% Align table columns on decimal point
\usepackage{bm}% bold math
\usepackage{hyperref}% add hypertext capabilities
\usepackage{braket}
\usepackage{siunitx}
\usepackage[utf8]{inputenc}
\usepackage{color} 



\begin{document}
	
\title{Amplitude and frequency sensing of microwave fields with \\ a superconducting transmon qudit}

\author{M. Kristen}	
\email{maximilian.kristen@kit.edu}
\author{A. Schneider}	
\author{A. Stehli}	
\author{T. Wolz}
\affiliation{Institute of Physics, Karlsruhe Institute of Technology, 76131 Karlsruhe, Germany}
\author{S. Danilin}
\affiliation{James Watt School of Engineering, University of Glasgow, Glasgow G12 8LT, UK}
\author{H. S. Ku}
\author{J. Long}
\author{X. Wu}
\author{R. Lake}
\author{D. P. Pappas}
\affiliation{National Institute of Standards and Technology, Boulder, Colorado 80309, USA}
\author{A. V. Ustinov}	
\affiliation{Institute of Physics, Karlsruhe Institute of Technology, 76131 Karlsruhe, Germany}
\affiliation{Russian Quantum Center, National University of Science and Technology MISIS, 119049 Moscow, Russia}
\author{M. Weides}
\email{martin.weides@glasgow.ac.uk}	
\affiliation{Institute of Physics, Karlsruhe Institute of Technology, 76131 Karlsruhe, Germany}
\affiliation{James Watt School of Engineering, University of Glasgow, Glasgow G12 8LT, UK}

\date{\today}
%-----------------------------------------------------------------------------

\begin{abstract}
	Experiments with superconducting circuits require careful calibration of the applied pulses and fields over a large frequency range. This remains an ongoing challenge as commercial semiconductor electronics are not able to probe signals arriving at the chip due to its cryogenic environment. Here, we demonstrate how the on-chip amplitude and frequency of a microwave signal can be inferred from the ac Stark shifts of higher transmon levels. In our time-resolved measurements we employ Ramsey fringes, allowing us to detect the amplitude of the systems transfer function over a range of several hundreds of MHz with an energy sensitivity on the order of $10^{-4}$. Combined with similar measurements for the phase of the transfer function, our sensing method can facilitate pulse correction for high fidelity quantum gates in superconducting circuits. Additionally, the potential to characterize arbitrary microwave fields promotes applications in related areas of research, such as quantum optics or hybrid microwave systems including photonic, mechanical or magnonic subsystems.
\end{abstract}

%-----------------------------------------------------------------------------

\maketitle

%-----------------------------------------------------------------------------
\section*{Introduction}
Implementing a fault-tolerant quantum processor requires gate fidelities far exceeding a threshold of 99\% \cite{Wang2011,Fowler2012,Campbell2017,Neill2018}. In superconducting qubits, these gates are realized by on or near-resonant microwave pulses \cite{Chow2012}. However, on the way to the circuit, the shape of these pulses is distorted by multiple passive microwave components such as attenuators, circulators and wires. These distortions negatively affect the gate fidelities if they are not accounted for.

The collective response of all microwave components to an incident signal is described by the transfer function of the system. If the transfer function is known, digital signal processing techniques allow for full control over the shape of applied pulses. However, since superconducting circuits are embedded in a cryogenic environment operated at millikelvin temperatures, the transfer function from pulse source to sample is not accessible with conventional network analyzers. In the past, this problem has been tackled by different calibration methods, which are usually limited to specific pulse shapes \cite{Gustavsson2013} or systems \cite{Bylander2009}. While more general pulse optimization schemes have been proposed theoretically, they have yet to be implemented in a real quantum system \cite{Bukov2018,Fosel2018,Niu2019}. 

In recent years, the growing interest in quantum sensors \cite{Degen2017,Bal2012,Honigl-Decrinis2019} has facilitated a more direct approach, where the signal arriving at the circuit is probed directly. In particular, superconducting qubits have been successfully employed as photon sensors due to their high electrical dipole moment. While sensing based on a variety of physical phenomena such as the cross-Kerr effect \cite{Hoi2013}, occurrence of the Mollow triplet \cite{Joas2017} or electromagnetically induced transparency \cite{Abdumalikov2010} has been shown, these methods are limited to the discrete frequencies of the qubit transitions. An alternative approach operates a qubit as a vector network analyzer, but only works in the MHz regime \cite{Jerger2019}. Recently, Schneider \textit{et al.} demonstrated that the ac Stark effect in anharmonic multi-level quantum systems (qudit) can be used to detect on-chip microwave fields \cite{Schneider2018}. Here, signals over a range of more than one GHz were measured. When including higher levels \cite{Braumuller2015}, this sensor can simultaneously determine the amplitude and frequency of an unknown signal, promoting it as a useful tool for experiments in quantum optics \cite{Deppe2008, Astafiev2010, You2011} and quantum microwave photonics \cite{Bozyigit2011,Menzel2012,Dmitriev2017}, where in-situ frequency detection can be beneficial. However, the spectroscopic measurement techniques employed in these proof of principle experiments offer limited precision for reasonable data acquisition times.            

In this work, we investigate the potential of the type of sensor used in Ref.\@ \cite{Schneider2018} to characterize the microwave transmission from source to sample. We use a time-resolved measurement setup to boost the sensor performance by an order of magnitude. By applying a well known microwave signal, we probe the amplitude of the transfer function over a wide frequency range. Finally, we estimate the errors and limits of our sensing scheme and discuss the potential for further improvement.

\section*{Results}

\begin{figure*}[t]
	\includegraphics{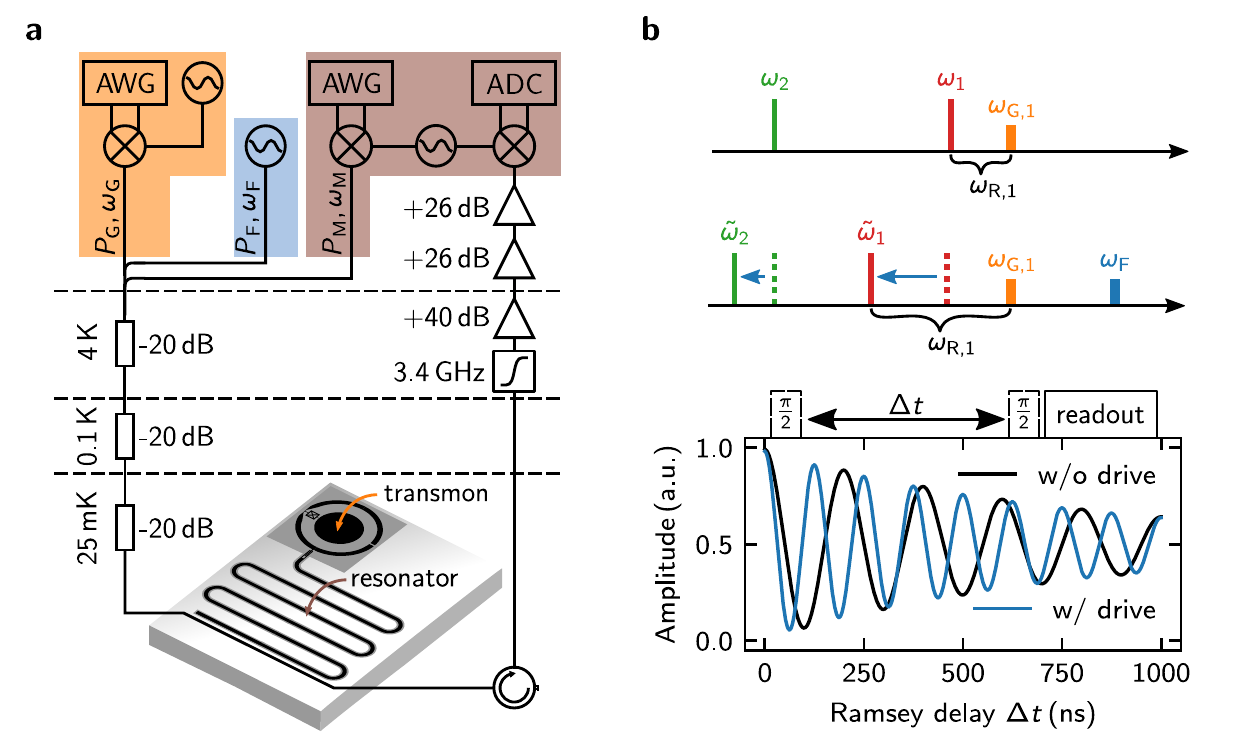}
	\caption{Experimental setup and methods. \textbf{a} Schematic diagram of the transmon qudit sensor and readout resonator (coplanar waveguide) connected to the employed microwave setup. The gate ($P_\text{G}, \omega_\text{G}$) and readout ($P_\text{M}, \omega_\text{M}$) pulses are merged with the continuous tone ($P_\text{F}, \omega_\text{F}$) which creates the on-chip microwave field to be measured. The combined signals are repeatedly attenuated within the cryostat before reaching the sample mounted at the $25\,\text{mK}$ stage. \textbf{b} Graphical representation of the sensor measurement procedure. As the qudit transitions $\omega_i$ ($i=1,2$) are shifted in the presence of a microwave field, the frequency $\omega_{\text{R},i}$ of the corresponding Ramsey oscillations changes, here exemplified for $\omega_1$. Ramsey fringes thus reveal the magnitude of these level shifts. Together, the shifts of the first two qudit levels can be used to extract the amplitude and frequency of the microwave field.}    
	\label{Fig1}
\end{figure*}

The sensor we use in our experiments is a non-tunable superconducting transmon ($\omega_1/2\pi = 4.685\,\text{GHz}$) with a concentric design \cite{Braumuller2016}.The transmon architecture offers a low anharmonicity ($280\,\text{MHz}$), which is beneficial for probing higher qudit transitions, as well as an enhanced dipole moment, which increases the sensitivity to local ac fields \cite{Koch2007}. To allow for manipulation ($P_\text{G}, \omega_\text{G}$) and readout (via a resonator) of the qudit, the sample is connected to a time resolved measurement setup (Fig.~\ref{Fig1}a). An additional microwave source with frequency $\omega_\text{F}$ and power $P_\text{F}$ was installed to generate a on-chip field with amplitude $A_\text{F} \propto \sqrt{10^{P_\text{F}/10\,\text{dBm}}}$. Neglecting the readout resonator, the Hamiltonian describing our system reads
\begin{eqnarray}
\label{Eq0}
H/\hbar &=& \sum_i \frac{E_{i}}{\hbar} \ket{i}\bra{i} + A_\text{G}(t)(\hat{b}+\hat{b}^\dagger)\cos\omega_\text{G}t \nonumber \\ && + A_\text{F}(\hat{b}+\hat{b}^\dagger)\cos\omega_\text{F}t,
\end{eqnarray}
where the anharmonic annihilation and creation operators $\hat{b}$ and $\hat{b}^\dagger$ take the different coupling strengths to the transmon levels into account, which are expressed in their Eigenbasis $\ket{i}$. The Eigenenergies $E_{i}$ are calculated from the exact solution of the Transmon Hamiltonian \cite{Koch2007}. In the following, we label the qudit transitions $\omega_i = E_i - E_{i-1}$ and their associated parameters with identical indices.  

To detect the amplitude and frequency of an on-chip microwave field we determine the ac Stark shift $\Delta_{i}$ that it induces in the first and second qudit transition ($i=1,2$). A simple but precise way to measure those shifts are Ramsey fringes \cite{Lee2002,Taylor2008}. The overall idea of the measurement scheme is sketched in Fig.~\ref{Fig1}b. Generally, performing Ramsey interferometry for a specific transition produces oscillations in the population of the associated qudit states. In the absence of an external field, the frequency of these oscillations simply depends on the frequency mismatch between the respective qudit transition and the applied gate tone $\omega_{\text{G},i}$. However, if the qudit is subjected to a microwave field, this mismatch changes due to the ac Stark effect. The shift of any qudit transition 
\begin{eqnarray}
\label{Eq1}
\Delta_i = \omega_{\text{R},i} - (\omega_{\text{G},i} - \omega_i) ,
\end{eqnarray}
can then be calculated from the oscillation frequency $\omega_{\text{R},i}$ corresponding to the respective Ramsey fringes, as long as the unperturbed qudit frequencies $\omega_i$ are known.

\begin{figure*}[t]
	\includegraphics[width=\textwidth]{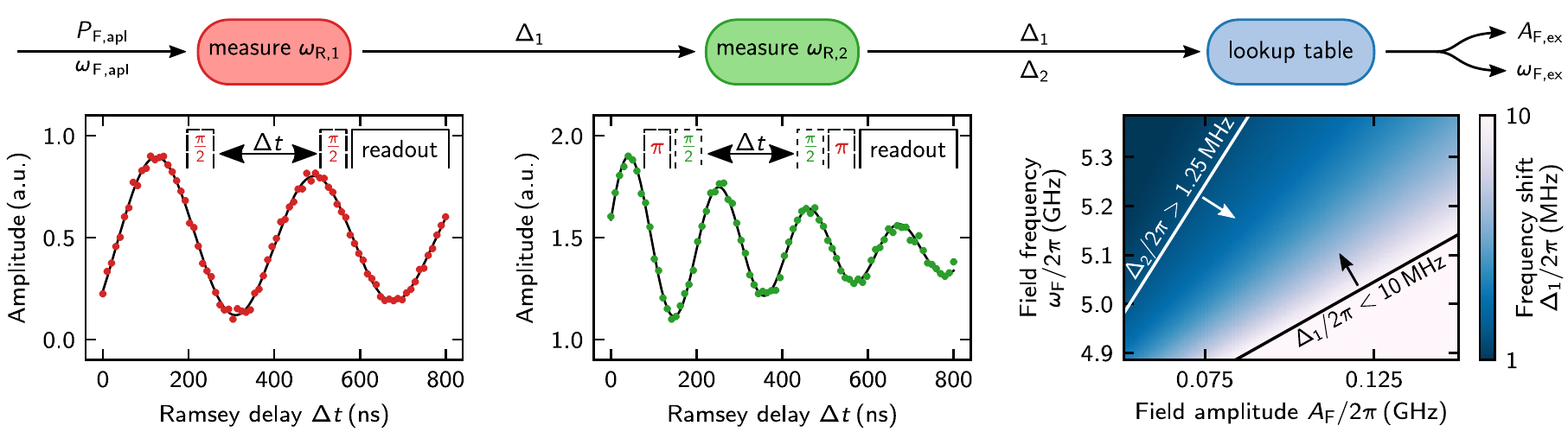}
	\caption{Employed sensing scheme. First, the shift of the first qudit transition $\Delta_{1}$ is determined from the frequency of the corresponding Ramsey oscillations. Second, $\Delta_{1}$ is used to adjust the resonant $\pi$-pulse which excites the qudit to the $\ket{1}$ state. Then, an additional Ramsey experiment is performed between $\ket{1}$ and $\ket{2}$ measuring $\Delta_{2}$. Third, $\Delta_{1}$ and $\Delta_{2}$ are processed by a pair of lookup tables to determine the frequency and amplitude of the microwave field causing the shifts. Here, only the lookup table for $\Delta_1$ is shown. The sensor limits depicted within the lookup table are derived in the main section.}
	\label{Fig2}
\end{figure*}

Figure~\ref{Fig2} shows Ramsey oscillations of the first and second qudit transition when applying a field with $\omega_\text{F,apl}/2\pi=5.285 \,\text{GHz}$ and $P_\text{F,apl}=4 \,\text{dBm}$. In the experiments, a $\pi$-pulse prior to the Ramsey sequence allows probing the frequency shift of the second excited state. An identical $\pi$-pulse after the sequence increases the visibility and removes the spurious signal of the relaxation to the ground state. For these $\pi$-pulses to be on resonance with the shifted transition frequency $\tilde{\omega}_1 = \omega_1 - \Delta_1$, knowledge of $\Delta_1$ is required. Consequently, the order in which the qudit transitions are probed is fixed. To determine the frequencies of the Ramsey oscillations, we fit the data with an exponentially damped sine function, which also accounts for the additional decay channels of higher lying qudit levels \cite{Peterer2015} via a declining amplitude offset. This decay of the higher excited level also limits the maximum Ramsey delay time $\Delta t$ used in our experiments (see Supplementary Information for details).

Lacking a closed analytical solution, the ac Stark shifts $\Delta_i$ calculated from $\omega_{\text{R},i}$ are then evaluated with a pair of lookup tables. Each lookup table contains the expected shifts of the respective qudit transition for various microwave fields. Searching both lookup tables simultaneously for the entries that are closest to our measurement data yields an unambiguous result for the frequency and amplitude of the detected field. In Ref.\@ \cite{Schneider2018}, these lookup tables are generated analytically by modeling the transmon as an anharmonic oscillator. The field dependent level shifts are then calculated from perturbation theory. However, we find that this simplified model is no longer accurate when detecting frequency shifts with a precision of a few kilohertz. We therefore rely on numerical simulations of the exact transmon Hamiltonian (Eq.~\eqref{Eq0}, see Methods for details).

The last plot in Fig.~\ref{Fig2} shows the numerically generated lookup table for the first qudit transition, illustrating the dependency of the ac Stark shift on the amplitude and frequency of the microwave field. Here, a black and white line represent the upper and lower limit of the sensor, respectively. These limits originate from the restricted number of measurement points for the Ramsey fringes and will be discussed in detail later. Evaluating the data in Fig.~\ref{Fig2} we find microwave photons of frequency $\omega_\text{F,ex}/2\pi=5.297 \,\text{GHz}$ arriving at the qudit at a rate of $A_\text{F,ex}/2\pi=0.097 \,\text{GHz}$, which corresponds to a power of $P_\text{F,ex} = A_\text{F,ex}\hbar\omega_\text{F,ex} = -116.7 \,\text{dBm}$.

\begin{figure*}[t]
	\includegraphics[width=\textwidth]{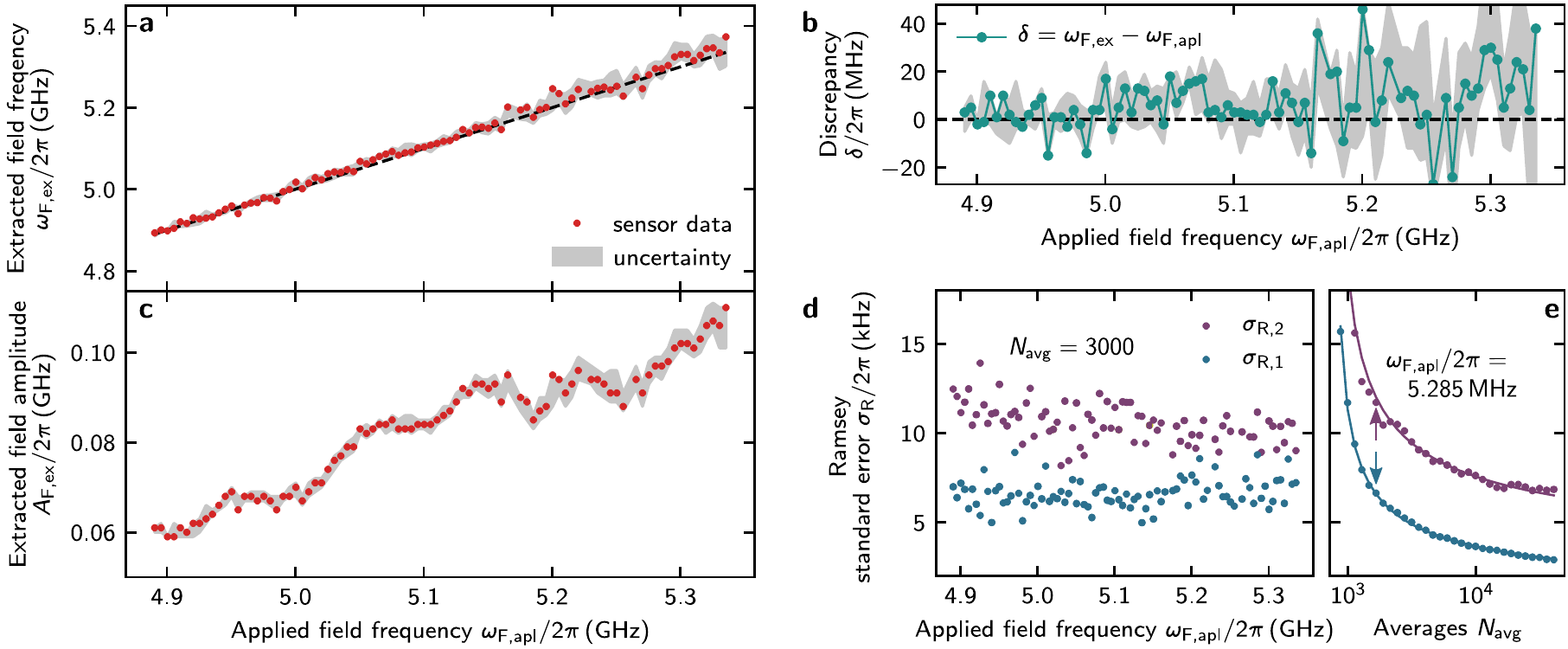}
	\caption{Sensor performance analysis. \textbf{a} Comparison between the frequencies applied with the microwave source ($\omega_\text{F,apl}$) and the frequencies extracted from the sensor ($\omega_\text{F,ex}$). The shaded area indicates the uncertainty estimated from the standard errors  to the Ramsey fits. \textbf{b} The magnitude of the discrepancy between $\omega_\text{F,apl}$ and $\omega_\text{F,ex}$ is an indicator for the reliability of our measurements. \textbf{c} Amplitude of the transfer function for a signal with constant power $P_\text{F,apl} = 4\,\text{dBm}$. \textbf{d} Ramsey standard errors used for the calculation of the uncertainty in (\textbf{a}) and (\textbf{c}). The values are extracted from the same fits as the sensor data. \textbf{e} Standard errors as a function of the number of averages. For this experiment, $N_\text{avg}=3000$ averages were used (indicated by the arrows).}
	\label{Fig3}
\end{figure*}

The full sensing scheme proposed in this work can be summarized as a three step process. After measuring the shift of the first and second qudit transition using Ramsey fringes, the field parameters are extracted from the measurement data with the help of pre-calculated lookup tables. To verify the scheme, we apply a well known microwave signal with constant power and gradually increase the frequency over a range of $450\,\text{MHz}$. We probe the field arriving on-chip with our sensor and plot the extracted $\omega_\text{F,ex}$ over the applied frequencies $\omega_\text{F,apl}$ (Fig.~\ref{Fig3}a), finding a good agreement. Plotting $A_\text{F,ex}$ over the same axis yields the amplitude of the transfer function (Fig.~\ref{Fig3}c). Here, we observe a strong frequency dependence, dominated by the readout resonator operating as a filter and cable resonances, which demonstrates the significance of calibrating microwave lines.   

Shaded areas in Fig.~\ref{Fig3} illustrate the uncertainty of our results. The uncertainty is estimated from varying our experimentally determined value of $\Delta_{i}$ by $\pm\sigma_{\text{R},i}$, where $\sigma_{\text{R},i}$ is the standard error of $\omega_{\text{R},i}$ resulting from the fit. In our case, $\sigma_{\text{R},i}$ is a consequence of the limited signal to noise ratio (SNR) during the measurement of individual data points and therefore depends on the number of averages $N_\text{avg}$ used in the experiments. As shown in Fig.~\ref{Fig3}e, the experimentally measured decline of $\sigma_{\text{R},i}$ is well fitted by ${\rm a}_i/\sqrt{N_\text{avg}}+{\rm c}_i$ (see also Supplementary Information), as expected from the shot noise limit \cite{Blanter2000}. In the interest of keeping the measurement time comparable to Ref.\@ \cite{Schneider2018}, all experiments were performed at $N_\text{avg}=3000$, fixing the errors at around $\sigma/2\pi = 10\,\text{kHz}$, see Fig.~\ref{Fig3}d. On average, this amounts to a relative uncertainty for the amplitude and frequency of $\Delta A_\text{F}/A_\text{F}=4\,\%$ and $\Delta\omega_\text{F}/\omega_\text{F}=0.5\,\%$, respectively. This error increases for higher frequencies, as $\Delta_i$ decreases for large detuning between microwave field and qudit, while the magnitude of $\sigma_{\text{R},i}$ remains unchanged.

Another potential source of noise, which has not been considered in the calculation above, are temporal fluctuations of the qudit transition frequencies $\Delta_i$ due to unstable two-level systems (TLS) \cite{Lisenfeld2015,Klimov2018, Schlor2019, Burnett2019}.  To quantitatively estimate their influence, we theoretically study the following example, where the first transition frequency is shifting by $\Delta\omega_1 = 20 \, \text{kHz}$ \cite{Schlor2019} right before a sensor measurement. Then, subsequent $\pi/2$-pulses are even further detuned and the corresponding Ramsey frequency will be altered, resulting in an offset for $\Delta_{1}$ by $\pm \Delta\omega_1$. Processing this offset together with the presented measurement data, we find that this causes an uncertainty for the extracted frequencies of $\overline{\Delta}\omega_\text{F}/2\pi=16.8 \, \text{MHz}$. Note that this uncertainty is independent from our evaluation of $\sigma_{\text{R},i}$, as the shift of the transition frequency affects all data points equally. This rough estimation thus provides a reasonable explanation for the few data points, where the discrepancies $\delta$ between $\omega_\text{F,ex}$ and $\omega_\text{F,apl}$ exceeding the estimated error bars in Fig.~\ref{Fig3}b. While a more profound analysis of this effect is challenging due to the varying timescales on which these fluctuations can occur, their influence could be mitigated in future measurements by a continuous recalibration of the qudit transition frequencies, i.e, adjusting the drive frequency to the fluctuating qudit transition frequencies.

In the following, we address the limits of our sensor (see Supplementary Information for an extended analysis). As discussed in Ref.\@ \cite{Schneider2018}, it is practical to limit the ac Stark qudit sensor to fields that are higher in frequency than the first qudit transition. Otherwise, the microwave field is more likely to excite higher qudit states. In this work, using Ramsey fringes results in additional constrains for the range of the sensor. The three parameters defining the total measurement time for a Ramsey experiment are the maximum delay time between the $\pi/2$-pulses $\Delta t_\text{max}$, the number of time steps $N_\text{R}$ and the passive reset time $T_\text{rep}$. To reduce the measurement time together with the chance of encountering frequency fluctuation \cite{Schlor2019}, it is desirable to minimize these parameters. At the same time, the sampling rate $f = N_\text{R}/\Delta t_\text{max}$ should be large enough to resolve the Ramsey oscillations clearly. Here, we find that values more than five times larger than the minimum value stated by the Nyquist-Shannon theorem \cite{Nyquist1928,Shannon1949} yield accurate fits. To ensure correct fitting of the data, it is also desirable to represent at least one full oscillation period within the measurement interval, which requires a sufficiently large $\Delta t_\text{max}$.

When operating the sensor with gate pulses that are on resonance with the unperturbed qudit frequency, Eq.~\eqref{Eq1} simplifies to $\Delta_{i}=\omega_{\text{R},i}$ and we can write the limits for the detectable frequency shifts as
\begin{eqnarray}
\label{Eq3}
\Delta_{1}/2\pi &<& N_\text{R}/(5 \cdot 2\Delta t_\text{max}) = 10 \, \text{MHz} \nonumber \\
\Delta_{2}/2\pi    &>& 1/(\Delta t_\text{max}) = 1.25 \, \text{MHz},
\end{eqnarray}
for $N_\text{R}=80$ and $\Delta t_\text{max}=800\,\text{ns}$. Together with $T_\text{rep}=\SI{240}{\micro\second}$ and $N_\text{avg}=3000$, all parameters amount to a total measurement time of $\sim \SI{1}{\minute}$. Note that the lower limit in Eq.~\eqref{Eq3} is given by $\Delta_2$, which is always a stronger constraint than $\Delta_1$. The lookup table in Fig.~\ref{Fig2} visualizes the set of detectable microwave fields determined by these limits. When a different range is required, they can be adjusted by choosing $\omega_{\text{G},i} \neq \omega_i$ or by changing the Ramsey parameters.

\section*{Discussion}
We have successfully implemented a sensor for microwave fields based on time-resolved measurements of the ac Stark shift. Employing Ramsey fringes, we harness the high sensitivity of the qudit phase on the frequencies of the first and second qudit transition. Evaluating the measured shifts with numerically generated lookup tables yields the amplitude and frequency of the applied microwave field. Using this sensing scheme, we measure the amplitude of the transfer function over a range of several hundreds of MHz. The results were validated by comparing the frequencies of the applied microwave tone with the sensor output. In comparison to the previous implementation by Schneider \textit{et al.} \cite{Schneider2018}, we were able to increase the precision by an order of magnitude to $\overline{\Delta}A_\text{F}/2\pi=3.4 \, \text{MHz}$ and $\overline{\Delta}\omega_\text{F}/2\pi=25 \, \text{MHz}$ for comparable measurement times. While a full pulse calibration requires similar measurements for the phase of the transfer function (see Supplementary Information for theoretical considerations), our results may already prove useful for advancing the control over hybrid microwave systems \cite{Xiang2013} and could enable broadband microwave detection in superconducting particle detectors \cite{Booth1996,Shokair2014}.

In the future, employing parametric amplifiers \cite{Castellanos-Beltran2008, Zhou2014, Winkel2020} and active reset \cite{Egger2018,Magnard2018,Gebauer2019} could reduce the measurement time of the sensor to a few seconds while simultaneously improving the precision. Moreover, advanced quantum sensing protocols that use linear slope detection over an extended dynamic range can be used to further increase the precision \cite{Berry2009,Cappellaro2012,Danilin2018}.

\section*{Methods}
\subsection*{Experimental setup}
We use a standard cQED setup consisting of a transmon qudit ($\omega_1/2\pi = 4.685\,\text{GHz}$ and $\omega_2/2\pi = 4.405\,\text{GHz}$) capacitively coupled to a $\lambda/2$-wavelength coplanar waveguide resonator ($\omega_\text{r}/2\pi = 6.878\,\text{GHz}$). To fabricate the resonator and the large-scale components of the transmon, thin-film NbTiN is used, whereas the Josephson tunnel junction consists of a conventional Al/AlO$_x$/Al stack \cite{Wu2017}. The chip is placed in a copper sample box and cooled down to temperatures below 25mK in a wet dilution refrigerator. 

The microwave gate pulses for the Ramsey sequence are generated in a single-sideband mixing scheme, using local oscillators and arbitrary waveform generators (AWG). Combined with the permanent microwave tone generating the on-chip field, these pulses are repeatedly attenuated on different temperature stages of the cryostat before reaching the sample. We use the resonator to dispersively readout the state of the qudit \cite{Bianchetti2009, Wallraff2004}. The readout signal is downconverted, digitized, and interpreted by our measurement software {\tt (\href{https://www.github.com/qkitgroup/qkit}{git.io/qkit})}.

\subsection*{Lookup table calculations}
Based on the full system Hamiltonian in Eq.~\eqref{Eq0}, we perform master-equation simulations using the QuTip package \cite{Johansson2012,Johansson2013}. Starting with the transmon in the ground (excited) state $\ket{0}$ ($\ket{1}$), we compute the full time evolution while applying a Ramsey sequence by temporarily switching on $A_\text{G}(t)$ in the simulation. After computing each point of the Ramsey fringes, $\Delta_{1}$ ($\Delta_{2}$) is determined by fitting the oscillations. This process is repeated for varying field amplitudes $A_\text{F}$ and frequencies $\omega_\text{F}$, gradually filling the lookup table. 

\section*{Data availability}
The data that support the findings of this study is available from the corresponding author upon reasonable request.

\section*{Code availability}
All data acquisition and analysis are performed with the open source measurement suite qkit {\tt (\href{https://www.github.com/qkitgroup/qkit}{git.io/qkit})}. The code used for the generation of the lookup tables is available from the corresponding author upon reasonable request.

\section*{Acknowledgments}
The authors thank J. Brehm, S. Schlör, J. I. Park and M. Vissers for their feedback and helpful discussions. This work was supported by the European  Research  Council  (ERC) under the Grant Agreement  No.\@  648011,  Deutsche Forschungsgemeinschaft (DFG)  projects INST 121384/138-1FUGG and WE 4359-7,  the Initiative and Networking Fund of the Helmholtz Association, and the state of Baden-Württemberg through bwHPC.  A.Sch. acknowledges financial support by the Carl-Zeiss-Foundation, A. St. by the Landesgraduiertenförderung (LGF) of the federal state Baden-Württemberg and T.W. by the Helmholtz International Research School for Teratronics (HIRST).  A.V.U.  acknowledges partial support from the Ministry of  Education and Science of the Russian Federation in the framework of  Contracts  No.\@  K2-2016-063 and No.\@ K2-2017-081.

\section*{Author contributions}
M.K. performed the measurements with support by A.Sch. ,A.St. and T.W.. H.S.K, J.L. and D.P.P. designed and X.W. and R.L. fabricated the sample. M.K. carried out the data analysis with contributions from A.Sch. and S.D. M.K. wrote the manuscript with input from all co-authors. A.V.U and M.W supervised the project. 

\section*{Competing interests}
The authors declare no competing interests.

%-----------------------------------------------------------------------------
\pagebreak
\onecolumngrid
\vspace{2cm}

\begin{center}
	\textbf{\Large Supplementary Information}
\end{center}

\section{Additional Sensor Limitations}         

In the main text we report how the parameter defining the Ramsey measurement (maximum delay time $\Delta t_\text{max}$, number of time steps $N_\text{R}$, number of averages $N_\text{avg}$) affect the range and precision of our sensor in practice. Here, we discuss additional constrains for the sensor performance as well as the ultimate limits of the sensing scheme.

\subsection{Qudit Coherence Time}
For our sensor, extremely weak or far detuned microwave fields result in very small frequency shifts, thus requiring long Ramsey sequences, i.e., large $\Delta t_\text{max}$, to be resolved. However, the Ramsey fringes will eventually be suppressed by the decoherence of the quantum state. By limiting $\Delta t_\text{max}$, this decoherence thus limits our sensor. Histogramms showing measured coherence times $T_2^i$ of the first and second qudit transition are depicted in Fig.~\ref{Fig1_sup}a. Due to the reduced coherence of higher transmon levels \cite{ Peterer2015}, $T_2^2$ sets the ultimate limit of the sensor. In our experiments, we therefore limit our maximum delay time to $\Delta t_\text{max} < 2\cdot T_2^2$. We note that larger coherence times ($T_2^1$) have been demonstrated with qudits of the same architecture \cite{Braumuller2016, Wu2017}. Unfortunately, our sample was limited by slotline modes.

\subsection{Sampling Rate}
In the opposite limit, where strong fields in close frequency range result in rather large transition frequency shifts and therefore fast Ramsey oscillations, the largest detectable frequency shift is proportional to  the number of time steps $N_\text{R}$. While $N_\text{R}$ is eventually limited by the sampling rate of the waveform generator, it is always possible to add a frequency offset to the Ramsey $\pi/2$-pulses to reduce the oscillation frequency and thereby shift the sensor limits.

\subsection{Ramsey Noise Floor}
In the main text we show that, in the absence of large qudit frequency fluctuations, the standard error of the Ramsey fit $\sigma_{\text{R},i}$ steadily declines with the number of used averages $N_\text{avg}$. The respective data is fitted by $\sigma_{\text{R},i}/2\pi=a_i/\sqrt{N_\text{avg}}+c_i$, where we include a constant offset that is independent of the averaging. Here, our fit yields
\begin{eqnarray}
\label{Eq1s}
c_{\rm 1} &=& 0.12 \, {\rm kHz}, \\
c_{\rm 2} &=& 2.20 \, {\rm kHz}
\end{eqnarray} 
for the first two transitions, respectively. This offset can be thought of as the noise floor of our sensor constituting an ultimate limit, below which no accurate information can be gained. While the details of its origin require further investigation, a possible explanation is low frequency 1/f noise from two level fluctuators \cite{Shnirman2002}. Due to the enhanced coupling matrix element of the second qudit transition \cite{Koch2007}, its noise floor is significantly higher. Moreover, probing the second transition requires more gates (see scheme in the main text). Thus, additional gate errors are introduced, also contributing to an increase of $c_{\rm 2}$. 	

\subsection{Readout Resonator}
Above the noise floor ($N_\text{avg}<10^5$), the precision of our sensor depends mainly on the signal to noise ratio (SNR) of our readout signal, which can quantitatively be written as $\text{SNR}_i = 1/\sigma_{\text{R},i} \approx \sqrt{N_\text{avg}}/a_i$ (with $a_{\rm 1} = 355 \, {\rm kHz}, \: a_{\rm 2} = 537 \, {\rm kHz}$). In the dispersive limit, the prefactor $a_i$ represents the ability to distinguish the frequencies of the resonator for different qudit states \cite{Blais2004}. This ability usually depends on both, the ratio between the dispersive shift $\chi$ and the full width at half maximum $\kappa$ of the resonator (see Fig.~\ref{Fig1_sup}b), as well as the ratio between the power of the readout signal and the thermal noise added by the amplifiers. 

\subsection{Background Microwave Power}
Spurious shifts caused by thermal photon population in the readout resonator ($w_{\rm r} = 6.878 \, {\rm GHz}$) are a well known noise source in cQED systems. For our setup, assuming a on-chip temperature of $75\, {\rm mK}$, Bose-Einstein statistics yields a mean photon number of $\overline{n} = 0.985$, which manifests as a constant offset of $\overline{n} \cdot \chi = 230.4 \, {\rm kHz}$ to the qudit frequency. While the offset will also slightly fluctuate due to the discrete nature of $\overline{n}$, these fluctuations only happen with respect to individual data points and are thus included in the error estimation presented in the main text. 

\subsection{Simulation Uncertainty}
\begin{figure*}[t]
	\includegraphics[width=\textwidth]{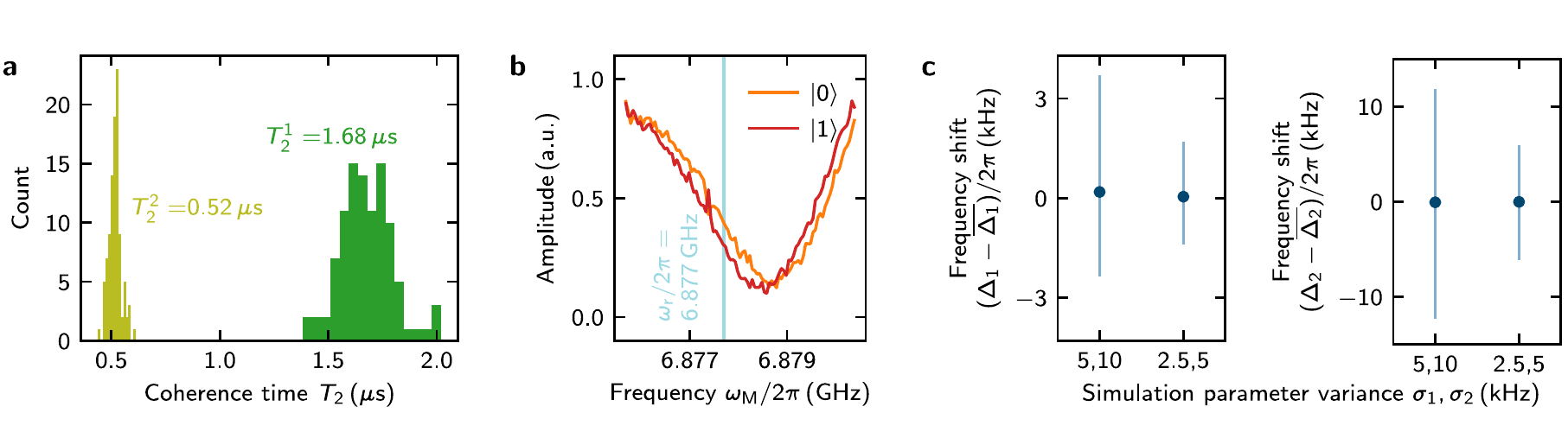}
	\caption{Additional sensor limitations. \textbf{a} Histograms showing the distribution of individual $T_2$ measurements for the first and second qudit transition. \textbf{b} Dispersive shift of our readout resonator ($\chi/\kappa = 0.08$). The separation between $\ket{0}$ and $\ket{1}$, which is one indicator for the SNR, is at maximum for the readout frequency $\omega_{\rm r}$. \textbf{c} Difference between simulation results using the measured qudit frequency and the mean value of all results obtained when varying these frequencies in the simulations by $\pm \sigma_{\text{R},1} ( \sigma_{\text{R},2})$. Error bars indicate the maximum spread of all simulation results.}    
	\label{Fig1_sup}
\end{figure*}

While our numerical simulations are in theory arbitrarily precise when taking enough qudit levels into account, they are still limited by the precision of the input parameter, i.e., the frequencies of the first and second qudit transition ($\omega_1, \omega_2$). To quantitatively estimate this effect, we perform numerical simulations of the expected frequency shifts for a fixed set of drive parameters ($\omega_\text{F,apl}/2\pi=4.985 \,\text{GHz}$, $A_\text{F,apl}/2\pi=0.07 \,\text{GHz}$) while varying $\omega_1$($\omega_2$) by $\pm \sigma_1 ( \sigma_2)$. The results are depicted in Fig.~\ref{Fig1_sup}c, where error bars show the maximum deviation from the mean value. While the uncertainties of the full lookup table will vary at each entry due to different drive parameters, this example illustrates that the potential error made with the simulation is comparable to the uncertainties of the input parameters, which are on the same order as in the actual experiment. We therefore think our error estimation based on the Ramsey standard error $\sigma_{\text{R},i}$ adequately covers our sensor uncertainty, since the actual uncertainty at worst becomes $2\sigma_{\text{R},i}$, for the unlikely case where the measurement and the simulation error fully add up. 

\subsection{High drive powers}  
It is well known that microwave fields with very high powers, both off and on resonance, can eventually excite higher levels in transmons \cite{Braumuller2015, Lescanne2019}. Consequently, our sensing method is limited to microwave field powers below a critical point where this effect dominates the system. While a general formulation for this mechanism is beyond the scope of this work, the problem can be adequately addressed by numerical simulations \cite{Pietikainen2018}. Thus, to estimate the drive power limit for our sensor, we simulate the population of higher transmon levels for a $200 \, {\rm MHz}$ detuned microwave with increasing amplitude:
\begin{enumerate}
	\item For $A_\text{F,apl}/2\pi=0.15 \,\text{GHz}$, which is the highest power simulated for our lookup table and well below the highest power measured in our experiment, the average population of the $\ket{2}$ state due to the drive is merely $1.1 \, \%$. The drive can then simply be treated as an additional noise source that adds to $\sigma_{\text{R},i}$.   
	\item For $A_\text{F,apl}/2\pi=0.75 \,\text{GHz}$, this probability increases to $22.7 \, \%$, making a coherent manipulation of the qudit challenging. In this parameter range, it might be more reasonable to use the spectroscopic measurement scheme presented in Ref.\@ \cite{Schneider2018}.     
	\item For $A_\text{F,apl}/2\pi=1.5 \,\text{GHz}$, there is a $2.25 \, \%$ probability to excite the qudit beyond the Josephson potential (here: $\ket{6}$ state), where all nonlinearities are lost.
\end{enumerate}

\section{Phase Measurement}
In order to use our sensor results for the proper calibration of microwave manipulation pulses, the phase of the transfer function is required.  Here, we present theoretical considerations for a measurement scheme that could allow for phase detection with our sensor.
The scheme consist of two consecutive $\pi/2$-pulses (denoted a and b), where the first pulse is on resonance with the first qudit transition $\omega_\text{a} = \omega_1$ and the second pulse is slightly detuned from it $\omega_\text{b} = \omega_1 + \Delta_{\rm b}$. Applying the rotating wave approximation, the Hamiltonians of these pulses can be written in the rotating frame
\begin{eqnarray}
\label{Eq2s}
H_{\rm a} = \begin{pmatrix} 0 & \Omega_{\rm a}  \\ \Omega_{\rm a} & 0  \\ \end{pmatrix}, \quad 
H_{\rm b} = \begin{pmatrix} -\Delta_{\rm b} & \Omega_{\rm b}e^{i\phi}  \\ \Omega_{\rm b}e^{-i\phi} & \Delta_{\rm b}  \\ \end{pmatrix}, 
\end{eqnarray}
where $\Omega_{\rm a,b}$ is the Rabi frequency (amplitude) of the respective pulse. Note that pulses a and b have the same phase at room temperature, but due to the systems phase response varying with frequency, the pulses reach the qudit with a relative phase difference $\phi = \phi(\Delta_{\rm b})$. To demonstrate how this phase can become experimentally accessible, we introduce the generalized Rabi frequency for the detuned pulse $\tilde{\Omega}_{\rm b} = \sqrt{\Omega_{\rm b}^2 + \Delta_{\rm b}^2}$ and then choose our pulse parameters such that $\Omega_{\rm a}t = \tilde{\Omega}_{\rm b}t = \pi/2$. While the resonant $\pi/2$-pulses can easily be calibrated with a standard Rabi sequence, the proper pulse length for the detuned pulse can also be determined from the amplitude of the transfer function initially measured with our sensor. Taking these relations into account, the unitary time evolution operator can be simplified to
\begin{eqnarray}
\label{Eq3s}
U_{\rm a}(0,t) = \exp(-\frac{iH_{\rm a}t}{\hbar}) = \frac{1}{\sqrt{2}}\begin{pmatrix} 1 & -i  \\ -i & 1  \\ \end{pmatrix}
\end{eqnarray} 
and
\begin{eqnarray}
\label{Eq4s}
U_{\rm b}(t,2t) = \exp(-\frac{iH_{\rm b}t}{\hbar}) = \frac{1}{\sqrt{2}}\begin{pmatrix} 1+ i \frac{\Delta_{\rm b}}{\tilde{\Omega}_{\rm b}} & -i \frac{\Omega_{\rm b}}{\tilde{\Omega}_{\rm b}}e^{i\phi}  \\ -i \frac{\Omega_{\rm b}}{\tilde{\Omega}_{\rm b}}e^{-i\phi} & 1 - i \frac{\Delta_{\rm b}}{\tilde{\Omega}_{\rm b}} \\ \end{pmatrix}.
\end{eqnarray}    
Turning back to the laboratory frame, we can calculate the probability to find the qudit in the excited state
\begin{eqnarray}
\label{Eq5s}
p_1(2t)=|\bra{1}U_{\rm a}(0,t)U_{\rm b}(t,2t)\ket{0}|^2= \frac{1}{2}\left(1+ \frac{\Omega_{\rm b}}{\tilde{\Omega}_{\rm b}}\cos\phi + \frac{\Delta_{\rm b}\Omega_{\rm b}}{\tilde{\Omega}_{\rm b}^2}\sin\phi \right).
\end{eqnarray} 
One can see that the final state probability depends on the phase difference between the pulses. This way, the frequency dependent phase of the transfer function can be measured relative to $\omega_1$. However, since $p_1$ vanishes for large $\Delta_{\rm b}$, this method only allows probing the phase in a limited frequency window around $\omega_1$.  While for our Rabi frequency $\Omega_{\rm a}=30\, {\rm MHz}$ an achievable sensor range of $\sim 100 \, {\rm MHz}$ is realistic, this would still be sufficient to cover the spectral width of typical manipulation pulses with a length of $100 \, {\rm ns}$.    

\begin{figure*}[t]
	\includegraphics[width=\textwidth]{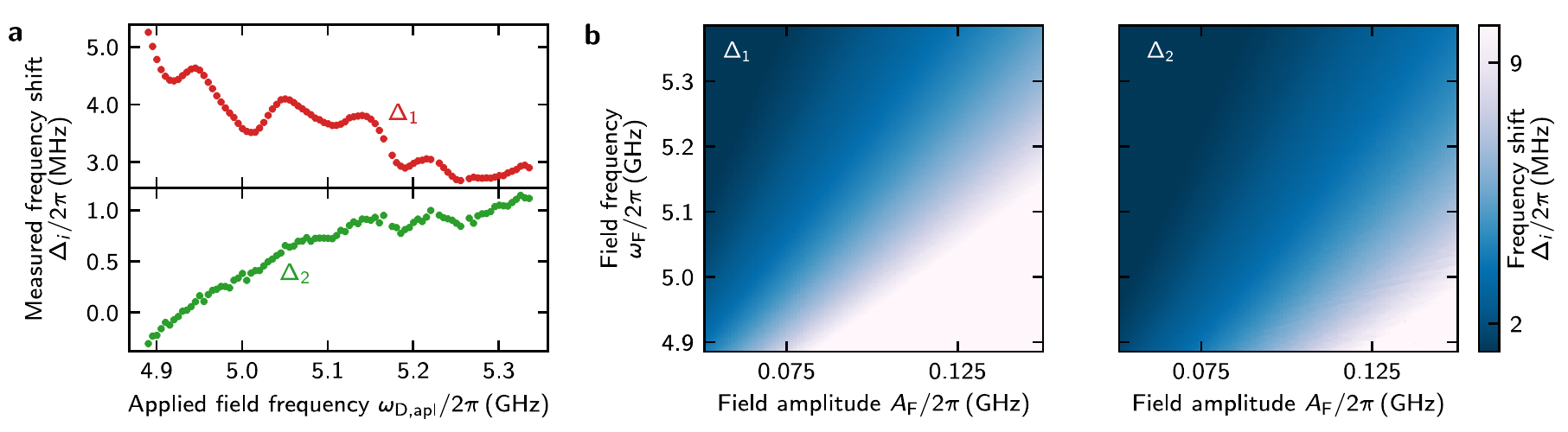}
	\caption{Unprocessed sensor data: \textbf{a} The measured frequency shifts and \textbf{b} the set of lookup tables used for their evaluation.}
	\label{Fig2_sup}
\end{figure*}

\section{Unprocessed sensor data and lookup tables} 
Figure~\ref{Fig2_sup} shows the measured frequency shifts (a) and the full set of lookup tables (b) that were used for the sensor performance analysis in the main text. From the unprocessed data, one can gain some interesting insights into the underlying sensor physics. Namely, that the qudit $\ket{1}$ and $\ket{2}$ state are nearly shifted in parallel, while the qudit ground state is not effected. As consequence, the variation of the field amplitude with frequency is much more profound in the first qudit transition and, more importantly, the second transition always shifts less than the first, which is why $\Delta_2$ is always a stronger constraint for the lower boundaries of the sensor. For further details, we refer the interested reader to Ref.\@ \cite{Schneider2018}.

\bibliographystyle{naturemag}
\bibliography{references_tot}

\end{document}